\documentstyle[preprint,aps,tabularx]{revtex}
\begin{document}
\preprint{\vbox{\hbox{IFP-790-UNC}
\hbox{hep-th/0011165}\hbox{November 2000} }}
\draft
\title{Finite N AdS/CFT Corespondence for Abelian and Nonabelian Orbifolds,
and Gauge Coupling Unification
\footnote{Contribution to the Journal of Mathematical Physics,
special issue devoted to Strings, Branes and M-theory.}}
\author{{\bf Paul H. Frampton}}
\address{Department of Physics and Astronomy,\\
University of North Carolina, Chapel Hill, NC 27599.}
\maketitle
\begin{abstract}
Although the AdS/CFT 
correspondence is rigorous only for
an infinite $N \rightarrow \infty$ stack of D3-branes, it
can be fruitfully studied for finite $N$ as a source
of gauge structures and choices for chiral fermions
and complex scalars which solve the hierarchy problem by
a conformal fixed point. We emphasize orbifolds $AdS_5 
\times S^5/\Gamma$
where the resulting GFT has ${\cal N} = 0$ 
supersymmetry.
The fact that the complex scalars are prescribed by the construction
limits the possible spontaneous symmetry breaking. 
Both abelian 
and nonabelian $\Gamma$ are illustrated by
simple examples. 
An accurate  
$sin^2 \theta$ in electroweak unification
can be obtained, suggesting that this approach
merits further study. 
\end{abstract}

\pacs{}

\bigskip

\newpage

\section{Introduction}

\bigskip
\bigskip

It has been a challenge over the last fifteen years to make a connection
between superstring theory and the real world.
The original attempts\cite{CHSW}
to identify massless string modes with the familiar degrees of freedom
(quarks, gluons, etc.) did not bear fruit
so other ideas to make such identification merit exploration.

A very old idea which is basic to string theory
is conformal invariance on the world sheet in two dimensions
\cite{P1,P2,BPZ,FQS,FMS}. A more recent idea is that conformal invariance
in four spacetime dimensions may
guide the sought-for connection
of superstring, and hence M theory, to
observable physics.

Let us briefly
outline the basis for AdS/CFT correspondence to set
the scene (for a more complete review, see \cite{AGMOO}).
Consider the type IIB superstring in flat ten-dimensional Minkowski
space, and a number $N$ of parallel
D3 branes close to each other, filling a $(3 + 1)$ subspace of
the $(9 + 1)$ spacetime.
The system has two types of perturbative excitations: closed and
open strings.
Closed strings are the excitations in the bulk and open strings
end on the D3 branes. At sufficiently
low energies ($\ll l_{string}^{-1}$, the string scale), only
massless states play a role.
The massless closed string states form a type IIB supergravity
multiplet: the
massless open string states form an ${\cal N} = 4$ vector supermultiplet
in $(3 + 1)$ dimensions with
interactions described by an ${\cal N} = 4$ $SU(N)$ supersymmetric gauge theory.

Next consider the same system but in the
background of a D3 brane solution of supergravity:
\begin{equation}
ds^2 = f^{-1/2} (-dt^2 + dx_1^2 + dx_2^2 + dx_3^2)
+ f^{1/2} (dr^2 + d\Omega_5)
\label{Dbrane}
\end{equation}
with
\begin{equation}
f = 1 + (R/r)^4 ~~~ {\rm and} ~~~ R = 4 \pi g_{string} (\alpha^{'}_{string})^2 N
\end{equation}

In this case, the energy of an object depends on $r$: if
it is $E_r$ at $r$ then at $r = \infty$ it appears redshifted
to $E_{\infty} = f^{-1/4} E_r$ because of the $g_{tt}$ metric component in
Eq.(\ref{Dbrane}).
Thus to an observer at infinity, the $r \rightarrow 0$ excitations 
in the "throat" appear of lowest energy.
There are two types of massless excitations. In the
bulk is the IIB supergravity multiplet 
interacting via supergravity: in the near-horizon region of the throat where
$r \ll R$ and $f \sim (R/r)^4$ the geometry is
that of $AdS_5 \times S^5$:
\begin{equation}
ds^2 = \left( \frac{r^2}{R^2} \right) (-dt^2 + dx_1^2 + dx_2^2 + dx_3^2) + \frac{R^2dr^2}{r^2}
+ R^2 d\Omega_5\
\label{AdS}
\end{equation}

In both backgrounds there are two decoupled theories in the low-energy limit.
One theory in both cases is supergravity in flat space. It is natural to identify the other two
theories: ${\cal N} = 4 ~~ SU(N)$ gauge theory
in $(3 + 1)$ spacetime corresponds to type IIB superstring theory
on $AdS_5 \times S^5$.

\bigskip
\bigskip

Let us now step back from M theory and address the needs of a theory
of the "real world". 

\bigskip

In particle phenomenology, 
the impressive success of the standard theory
based on $SU(3) \times SU(2) \times U(1)$ has naturally led to the question
of how to extend the theory to higher energies. One is necessarily led by
weaknesses and incompleteness in the standard theory. If one extrapolates
the standard theory as it stands one finds (approximate) unification of the
gauge couplings at $\sim 10^{16}$ GeV. But then there is the {\it hierarchy}
problem of how to explain the occurrence of the tiny dimensionless ratio 
$\sim 10^{-14}$ of the weak scale to the unification scale. Inclusion of
gravity leads to a {\it super-hierarchy} problem of the ratio of the weak
scale to the Planck scale, $\sim 10^{18}$ GeV, an even tinier $\sim 10^{-16}$.
Although this is obviously a very important problem about which
conformality by itself is not informative, we shall discuss first the
hierarchy rather than the super-hierarchy.

\bigskip

There are four well-defined approaches to the hierarchy problem:

\begin{itemize}
\item  1. Supersymmetry

\item  2. Technicolor.

\item  3. Extra dimensions.

\item  4. Conformality.
\end{itemize}

\noindent {\it Supersymmetry} has the advantage of rendering the hierarchy
technically natural, that once the hierarchy is put in to the lagrangian it
need not be retuned in perturbation theory. Supersymmetry predicts
superpartners of all the known particles and these are predicted to be at or
below a TeV scale if supersymmetry is related to the electroweak breaking.
Inclusion of such hypothetical states improves the gauge coupling
unification. On the negative side, supersymmetry does not explain the origin
of the hierarchy.

\bigskip

\noindent {\it Technicolor} postulates that the Higgs boson is a 
fermion-antifermion composite
bound by a new (technicolor) strong dynamics at or
below the TeV scale. This obviates the hierarchy problem. On the minus side,
no simple convincing model of technicolor has been found.

\bigskip

\noindent {\it Extra dimensions} can have a range as large as 
$1 ({\rm TeV})^{-1}$ and the 
gauge coupling unification can happen quite differently than
in only four spacetime dimensions. This replaces the hierarchy problem with
a different fine-tuning question of why the extra dimension is restricted to
a distance corresponding to the weak interaction scale.
There is also a potentially serious problem with the proton lifetime.

\bigskip

\noindent {\it Conformality} is inspired by 
the AdS/CFT correspondence discussed above
superstring duality and assumes
that the particle spectrum of the standard model is enriched such that there
is a conformal fixed point of the renormalization group at the TeV scale.
Above this scale the coupling do not run so the hierarchy is nullified.
Instead, the couplings $\alpha_1^{-1}, \alpha_2^{-1}, \alpha_3^{-1}$ run from low
energy up
to the TeV scale then combine to one
energy-independent $\alpha_{conformal}^{-1}$ (in most cases equal to
$\alpha_3^{-1}$).
It is important to realize that the
observed difference between $\alpha_1^{-1}$, $\alpha_2^{-1}$ and $\alpha_3^{-1}
= \alpha_{conformal}^{-1}$ arise in this approach from
the group theory associated with embedding 
$SU(3) \times SU(2) \times U(1)$ in
a semi-simple unifying gauge group.

\bigskip

Conformality is the approach followed in this paper. We shall systematically
analyse the compactification of the IIB superstring on $AdS_5 \times
S^5/\Gamma$ where $\Gamma$ is a discrete non-abelian group designed to
break all supersymmetries.

Until very recently, the possibility of testing string theory seemed remote,
at best. The advent of the $AdS/CFT$ correspondence
suggests this point of view may be too pessimistic, since it could lead to 
$\sim TeV$ evidence for strings. With this thought in mind, we are
encouraged to build $AdS/CFT$ models with realistic fermionic structure, and
reduce to the standard model below $\sim 1TeV$.

Using AdS/CFT duality, one arrives at a class of gauge field theories of
special recent interest. The simplest compactification of a ten-dimensional
superstring on a product of an AdS space with a five-dimensional spherical
manifold leads to an ${\cal N} = 4~SU(N)$ supersymmetric gauge theory, well
known to be conformally invariant\cite{mandelstam}. By replacing the
manifold $S^5$ by an orbifold $S^5/\Gamma$ one arrives at less
supersymmetries corresponding to ${\cal N} = 2,~1 ~{\rm or}~ 0$ depending
\cite{KS} on whether $\Gamma \subset SU(2), ~~ SU(3), ~~{\rm or} 
\not\subset SU(3)$ respectively, where $\Gamma$ is in all cases a subgroup of 
$SU(4) \sim SO(6)$ the isometry of the $S^5$ manifold.

It was conjectured in \cite{maldacena} that such $SU(N)$ gauge theories are
conformal in the $N \rightarrow \infty$ limit. In \cite{F1} it was
conjectured that at least a subset of the resultant nonsupersymmetric ${\cal 
N} = 0$ theories are conformal even for finite $N$ and that
one of this subset provides the right extension
of the standard model. Some first steps to
check this idea were made in \cite{WS}. Model-building based on abelian 
$\Gamma$ was studied further in \cite{CV,F2}, arriving in \cite{F3} at an 
$SU(3)^7$ model based on $\Gamma = Z_7$ which has three families of chiral
fermions, a correct value for ${\rm sin}^2 \theta$ and a conformal scale $\sim 10$~~TeV.

\newpage

\section{Abelian Orbifolds}

\bigskip
\bigskip
\bigskip

Since, in the context of field-string duality, there has been a shift
regarding the relationship of gravity
to the standard model of strong and electroweak interactions we
shall begin by characterising how gravity fits in,
then to suggest more
specifically how the standard model fits in to the string framework.

The descriptions of gravity and of the standard model are contained
in the string theory. In the string picture in ten spacetime dimensions,
or upon compactification to four dimensions, there is a massless spin-two
graviton but the standard model is not manifest in the way we
shall consider it. In the conformal field theory
extension of the standard model, gravity is strikingly absent.
The field-string duality does not imply that the standard model
already contains gravity and, in fact, it does not.

In the field theory description\cite{F1,WS,CV,F2}
used in this article, one will simply ignore the
massless spin-two graviton. Indeed,
since we are using the field theory description
only below the conformal scale of $\sim 1$TeV
( or, as suggested later in
this paper, 10TeV) and forgoing any requirement of grand unification,
the hierarchy between the weak scale and theory-generated
scales like $M_{GUT}$ or $M_{PLANCK}$ is resolved.
Moreover, seeking
the graviton in the field theory description is
possibly resolvable by going to a higher dimension
and restricting the range of the higher dimension.
Here we are looking only at the strong and weak interactions at accessible
energies below, say, 10TeV.

Of course, if we ask questions in a different regime, for example about
the scattering of particles with center-of-mass energy of the
order $M_{PLANCK}$, then the graviton
will become crucial\cite{Hooft,ACV} and a string, rather than a
field, description will be the viable one.

It is important to distinguish between the holographic
description of the five-dimensional gravity
in $(AdS)_5$ made by the four-dimensional CFT and the origin of the
four-dimensional graviton. The latter could be described holographically
only by a lower three-dimensional field theory which is not
relevant to the real world. Therefore the graviton
of our world can only arise by {\it compactification} of a higher
dimensional graviton.
Introduction of gravity must break conformal
invariance and it is an interesting
question whether
this breaking is related to the mass and
symmetry-breaking scales in the low-energy theory. That is
all I will say about gravity in the present paper; the
remainder is on the standard model and its embedding in a CFT.

An alternative to conformality, grand unification with supersymmetry,
leads to an impressively accurate gauge coupling unification\cite{ADF,ADFFL}.
In particular it predicts an electroweak mixing angle
at the Z-pole, ${\tt sin}^2 \theta = 0.231$. This result
may, however, be fortuitous, but rather than
abandon gauge coupling unification, we can rederive ${\tt sin}^2 \theta = 0.231$
in a different way by embedding the electroweak $SU(2) \times U(1)$ in
$SU(N) \times SU(N) \times SU(N)$ to
find ${\tt sin}^2 \theta = 3/13 \simeq 0.231$\cite{F2}.
This will be a common feature of the models in this paper.

Actually, it may be premature to dismiss as accidental the success of grand unification
with ${\tt sin}^2\theta$ since the principal topic here (${\tt AdS/CFT}$)
teaches us that quite different theoretical descriptions can be "dual"
and the same may eventually be understood for conformality and
grand unification. For example,
conformality is compatible with supersymmetry at low-energy.

The conformal theories will be finite without
quadratic or logarithmic divergences. This requires appropriate
equal numbers of fermions and bosons which can cancel in
loops
and which occur without the necessity of space-time supersymmetry.
As we shall see in one example, it is possible to combine
spacetime supersymmetry
with conformality but the latter is the driving principle and the former
is merely an option: additional fermions and scalars are predicted by
conformality in the TeV range\cite{CV,F2},
but in general these particles are different and distinguishable
from supersymmetric partners.
The boson-fermion cancellation is essential for the
cancellation of infinities, and will
play a central role in the calculation of the cosmological constant
(not discussed here). In the field picture, the
cosmological constant measures the vacuum energy density.

Here we shall focus on abelian orbifolds characterised by the discrete group
$Z_p$. Non-abelian orbifolds will be discussed in the next section.

The steps in building a model for the abelian case (parallel steps
hold for non-abelian orbifolds) are:

\begin{itemize}

\item{(1)} Choose the discrete group $\Gamma$. Here we are considering
only $\Gamma = Z_p$. We define $\alpha = {\rm exp}(2 \pi i/p)$.

\item{(2)} Choose the embedding of $\Gamma \subset SU(4)$ by
assigning ${\bf 4} = (\alpha^{A_1}, \alpha^{A_2}, \alpha^{A_3}, \alpha^{A_4})$
such that $\sum_{q=1}^{q=4} A_q = 0 ({\rm mod} p)$. To
break ${\cal N} = 4$ supersymmetry to
${\cal N} = 0$ ( or ${\cal N} = 1$) requires that
none (or one) of the $A_q$ is equal to zero (mod p).

\item{(3)} For chiral fermions one requires that ${\bf 4} \not\equiv {\bf 4^{*}}$
for the embedding of $\Gamma$ in $SU(4)$.

The chiral fermions are in the bifundamental representations of $SU(N)^p$
\begin{equation}
\sum_{i=1}^{i=p} \sum_{q=1}^{q=4}
(N_i, \bar{N}_{i + A_q})
\label{fermions}
\end{equation}
If $A_q=0$ we interpret $(N_i, \bar{N}_i)$ as a singlet plus an adjoint
of $SU(N)_i$.

\item{(4)} The {\bf 6} of $SU(4)$ is real {\bf 6} = $(a_1, a_2, a_3, -a_1, -a_2, -a_3)$
with
$a_1 = A_1 + A_2$,
$a_2 = A_2 + A_3$,
$a_3 = A_3 + A_1$
(recall that all components are defined modulo p).
The complex scalars are in the bifundamentals
\begin{equation}
\sum_{i=1}^{i=p} \sum_{j=1}^{j=3}
(N_i, \bar{N}_{i \pm a_j})
\label{scalars}
\end{equation}
\noindent The condition in terms of $a_j$ for ${\cal N} = 0$ is
$\sum_{j=1}^{j=3} (\pm a_j) \not= 0 ({\rm mod}~~ p)$\cite{F1}.
\item{(5)} Choose the $N$ of $\bigotimes_i SU(Nd_i)$ (where the $d_i$ are the
dimensions of the representrations of $\Gamma$). For the abelian case
where $d_i \equiv 1$, it is natural to choose $N=3$ the largest
$SU(N)$ of the standard model (SM) gauge group. For a non-abelian $\Gamma$
with $d_i \not\equiv 1$ the choice $N=2$ would be indicated.

\item{(6)}  The $p$ quiver nodes are identified as color (C), weak isospin (W),
or a third $SU(3)$ (H). This specifies the embedding of the gauge group
$SU(3)_C \times SU(3)_W \times SU(3)_H \subset \bigotimes SU(N)^p$.

This quiver node identification is guided by (7), (8) and (9) below.

\item{(7)}  The quiver node identification is required to
give three chiral families under Eq.(\ref{fermions})
It is sufficient to make three of the $(C + A_q)$ to be W and the fourth H, given that there is only
one C quiver node, so that there are three $(3, \bar{3}, 1)$. Provided that
$(\bar{3}, 3, 1)$ is avoided by the $(C - A_q)$ being H, the remainder
of the three family trinification will be automatic by chiral anomaly cancellation.
Actually, a sufficient condition for three families has been given; it is
necessary only that the difference between the number of $(3 + A_q)$
nodes and the number of $(3 - A_q)$ nodes
which are W be equal to three.

\item{(8)}  The complex scalars of Eq. (\ref{scalars}) must be sufficient for their
vacuum expectation values (VEVs) to spontaneously break
$SU(3)^p \longrightarrow SU(3)_C \times SU(3)_W \times SU(3)_H
\longrightarrow SU(3)_C \times SU(2)_W \times U(1)_Y \longrightarrow SU(3)_C \times U(1)_Q$.

Note that, unlike grand unified theories (GUTs) with or without supersymmetry,
the Higgs scalars are here prescribed by the conformality condition.
This is more satisfactory because it implies that the Higgs sector cannot be chosen
arbitrarily.

\item{(9)} Gauge coupling unification should apply at least to the electroweak mixing
angle ${\rm sin}^2 \theta = g_Y^2 / (g_2^2 + g_Y^2) \simeq 0.231$. For trinification
$Y = 3^{-1/2} ( - \lambda_{8W} + 2\lambda_{8H})$ so that $(3/5)^{1/2} Y$ is
correctly normalized. If we make $g_Y^2 = (3/5)g_1^2$ and $g_2^2 = 2 g_1^2$
then ${\rm sin}^2 \theta = 3/13 \simeq 0.231$ with sufficient accuracy.

\end{itemize}

\bigskip
\bigskip

In the remainder of this section we answer all these steps for the choice $\Gamma = Z_p$
for successive $p = 2, 3 ...$ up to $p = 7$, then add some concluding remarks.

\bigskip
\bigskip

\begin{itemize}

\item{{\bf p = 2}}

In this case $\alpha = -1$ and therefore one cannot
construct any complex {\bf 4} of $SU(4)$ with
${\bf 4} \not\equiv {\bf 4^*}$.
Chiral fermions are therefore
impossible.

\bigskip

\newpage

\bigskip

\item{{\bf p = 3}}

The only possibilities are $A_q = (1, 1, 1, 0)$ or $A_q = (1, 1, -1, -1)$.
The latter is real and leads to no chiral fermions.
The former leaves ${\cal N} = 1$ supersymmetry and is a simple three-family
model\cite{KS} by the quiver node identification C - W - H. The scalars $a_j = (1, 1, 1)$
are sufficient to spontaneously break to the SM. Gauge coupling unification is, however,
missing since ${\rm sin}^2 \theta = 3/8$, in
bad disagreement with experiment.

\item{{\bf p = 4}}

The only complex ${\cal N} = 0$ choice is $A_q = (1, 1, 1, 1)$. But
then $a_j = (2, 2, 2)$ and any quiver node identification
such as C - W - H - H has 4 families and the scalars are insufficient to
break spontaneously the symmetry to the SM gauge group.

\item{{\bf p = 5}}

The two inequivalent complex choices are $A_q = (1, 1, 1, 2)$ and $A_q = (1, 3, 3, 3)$.
By drawing the quiver, however, and using the rules
for three chiral families given in (7)
above, one finds that the node identification and the prescription of the scalars
as $a_j = (2, 2, 2)$ and $a_j = (1, 1, 1)$ respectively does not permit
spontaneous breaking to the standard model.

\item{{\bf p = 6}}

Here we can discuss three inequivalent complex possibilities as follows:

\noindent (6A) $A_q = (1, 1, 1, 3)$ which implies $a_j = (2, 2, 2)$.

\bigskip

\noindent Requiring three families means a node identification C - W - X - H - X - H
where X is either W or H. But whatever we choose for the X the scalar representations
are insufficient to break $SU(3)^6$ in the desired fashion down to the 
standard theory. This
illustrates the difficulty of model building when the scalars are not
in arbitrary representations.

\noindent (6B) $A_q = (1, 1, 2, 2)$ which implies $a_j = (2, 3, 3)$.

\bigskip

\noindent
Here the family number can be only zero, two or four as can be seen by inspection
of the $A_q$ and the related quiver diagram. So (6B) is of no phenomenological interest.

\noindent (6C) $A_q = (1, 3, 4, 4)$ which implies $a_j = (1, 1, 4)$.

\bigskip

\noindent Requiring three families needs a quiver node identification which is of the form
{\it either}
C - W - H - H - W - H {\it or} C - H - H - W - W - H. The scalar representations
implied by $a_j = (1, 1, 4)$ are, however, easily seen to be insufficient to do the required
spontaneous symmetry breaking (S.S.B.) for both of these identifications.

\item{{\bf p =7}}

Having been stymied mainly by the rigidity of the scalar representation
for all $p \leq 6$, for $p = 7$ there
are the first cases which work. Six inequivalent complex embeddings of $Z_7
\subset SU(4)$ require consideration.

\noindent (7A) $A_q = (1, 1, 1, 4) \Longrightarrow a_j = (2, 2, 2)$

\bigskip

For the required nodes C - W - X - H - H - X - H the
scalars are {\it insufficient} for S.S.B.

\bigskip

\noindent (7B) $A_q = (1, 1, 2, 3) \Longrightarrow a_j = (2, 3, 3)$

\bigskip

The node identification C - W - H - W - H - H - H leads
to a {\it successful} model.

\bigskip

\noindent (7C) $A_q = (1, 2, 2, 2) \Longrightarrow a_j = (3, 3, 3)$

\bigskip

Choosing C - H - W - X - X - H - H to derive three families, the
scalars {\it fail} in S.S.B.

\bigskip

\noindent (7D) $A_q = (1, 3, 5, 5) \Longrightarrow a_j = (1, 1, 3)$

\bigskip

The node choice C - W - H - H - H - W - H leads
to a {\it successful} model. This is Model A of \cite{F2}.

\bigskip

\noindent (7E) $A_q = (1, 4, 4, 5) \Longrightarrow a_j = (1, 2, 2)$

\bigskip

The nodes C - H - H - H - W - W - H are
{\it successful}.

\bigskip

\noindent (7F) $A_q = (2, 4, 4, 4) \Longrightarrow a_j = (1, 1, 1)$

\bigskip

Scalars {\it insufficient} for S.S.B.

\end{itemize}

The three successful models (7B), (7D) and (7E) lead to
an $\alpha_3(M) \simeq 0.07$. Since $\alpha_3(1 {\rm TeV}) \geq 0.10$
this suggest a conformal scale $M \simeq 10$ TeV \cite{F2}.
The above models have less generators than an
$E(6)$ GUT and thus $SU(3)^7$ merits
further study. It is possible, and under investigation, that non-abelian
orbifolds will lead to a simpler model.

\bigskip

For such field theories it is important to establish the existence of a fixed
manifold with respect to the renormalization group. It
could be a fixed line but more likely, in
the ${\cal N} = 0$ case, a fixed point. It is known that in the
$N \longrightarrow \infty$ limit the theories become conformal, but
although this 't Hooft limit\cite{tH} is where the field-string duality
is derived we know that finiteness survives to finite N in the
${\cal N} = 4$ case\cite{mandelstam} and this makes it plausible
that at least a conformal point occurs also for the ${\cal N} = 0$
theories with $N = 3$ derived above.

The conformal structure cannot by itself predict all the dimensionless ratios of
the standard model such as mass ratios and mixing angles
because these receive contributions,
in general, from soft breaking of conformality. With a
specific assumption about the pattern of conformal
symmetry breaking, however, more work should lead to
definite predictions for such quantities.

\bigskip
\bigskip

\newpage

\bigskip
\bigskip

\section{Nonabelian Orbifolds}

\bigskip
\bigskip

Abelian orbifolds lead us to consider the finite $N$ value
$N = 3$ guided by trinification $SU(3)^3$ and
the fact that all representations of abelian groups $Z_p$
are one dimensional. 

A nonabelian orbifold can allow the consideration of finite $N = 2$ since for
a $\Gamma$ with doublet and
singlet representations
can lead to a generalization of a left-right structure of the type:
$SU(4) \times SU(2) \times SU(2)$.

\bigskip
\bigskip

First we remind the reader of available nonabelian
$\Gamma$ of low order $g \leq 31$.

\bigskip
\bigskip

Of the nonabelian finite groups, the best known are perhaps the
permutation groups $S_N$ (with $N \geq 3$) of order $N!$
The smallest non-abelian finite group is $S_3$ ($\equiv D_3$),
the symmetry of an equilateral triangle with respect to all
rotations in a three dimensional sense. This group initiates two
infinite series, the $S_N$ and the $D_N$. Both have elementary
geometrical significance since the symmetric permutation group
$S_N$ is the symmetry of the N-plex in N dimensions while the dihedral group
$D_N$ is the symmetry of the planar N-agon in 3 dimensions.
As a family symmetry, the $S_N$ series becomes uninteresting rapidly
as the order and the dimensions of the representions increase. Only $S_3$
and $S_4$ are of any interest as symmetries associated with the particle
spectrum\cite{Pak}; also, the order (number of elements) of the $S_N$ groups
grow factorially with N. The order of the dihedral groups increases only
linearly with N and their irreducible representations are all one- and
two- dimensional. This is reminiscent of the representations of the
electroweak $SU(2)_L$ used in Nature.

Each $D_N$ is a subgroup of $O(3)$ and has a counterpart double dihedral
group $Q_{2N}$, of order $4N$, which is a subgroup of the double covering
$SU(2)$ of $O(3)$.

With only the use of $D_N$, $Q_{2N}$, $S_N$ and the tetrahedral group T (of
order
12, the even permutations subgroup of $S_4$ ) we find 32 of the 45
nonabelian groups up to order 31, either as simple groups or as
products of simple nonabelian groups with abelian groups:
(Note that $D_6 \simeq Z_2 \times D_3, D_{10} \simeq Z_2 \times D_5$ and $
D_{14} \simeq Z_2 \times D_7$ )

\begin{center}

\begin{tabular}{||c||c||}   \hline
g & \\    \hline
$6$  & $D_3 \equiv S_3$\\  \hline
$8$ & $ D_4 , Q = Q_4 $\\    \hline
$10$& $D_5$\\   \hline
$12$&  $D_6, Q_6, T$ \\ \hline
$14$& $D_7$\\  \hline
$16$& $D_8, Q_8, Z_2 \times D_4, Z_2 \times Q$\\  \hline
$18$& $D_9, Z_3 \times D_3$\\  \hline
$20$& $D_{10}, Q_{10}$ \\  \hline
$22$& $D_{11}$\\  \hline
$24$& $D_{12}, Q_{12}, Z_2 \times D_6, Z_2 \times Q_6, Z_2 \times T$,\\  \hline
 & $Z_3 \times D_4, Z_3 \times Q, Z_4 \times D_3, S_4$\\  \hline
$26$& $D_{13}$\\  \hline
$28$& $D_{14}, Q_{14}$ \\  \hline
$30$& $D_{15}, D_5 \times Z_3, D_3 \times Z_5$\\  \hline
\end{tabular}

\end{center}

\bigskip
\bigskip

There remain thirteen others formed by twisted products of abelian factors.
Only certain such twistings are permissable, namely (completing all $g \leq 31$).

$$\begin{tabular}{||c||c||}   \hline
g & \\    \hline
$16$  & $Z_2 \tilde{\times} Z_8$ (two, excluding $D_8$), $Z_4 \tilde{\times}
Z_4, Z_2 \tilde{\times}(Z_2 \times Z_4)$
(two)\\  \hline
$18$ & $Z_2 \tilde{\times} (Z_3 \times Z_3)$\\    \hline
$20$&  $Z_4 \tilde{\times} Z_5$ \\   \hline
$21$&  $Z_3 \tilde{\times} Z_7$ \\    \hline
$24$&  $Z_3 \tilde{\times} Q, Z_3 \tilde{\times} Z_8, Z_3 \tilde{\times} D_4$
\\  \hline
$27$&  $ Z_9 \tilde{\times} Z_3, Z_3 \tilde{\times} (Z_3 \times Z_3)$ \\
\hline
\end{tabular}$$

It can be shown that these thirteen exhaust the classification of {\it all}
inequivalent finite groups up to order thirty-one\cite{books,FK2}.

Of the 45 nonabelian groups, the dihedrals ($D_N$) and double dihedrals
($Q_{2N}$), of order 2N and 4N respectively,
form the simplest sequences. In particular, they fall into subgroups of
$O(3)$ and $SU(2)$ respectively,
the two simplest nonabelian continuous groups.

For the $D_N$ and $Q_{2N}$, the multiplication tables, as derivable from the
character tables,
are, in general, simple to express. $D_N$, for odd N, has two singlet
representations $1,1^{'}$ and $m = (N-1)/2$
doublets $2_{(j)}$ ($1 \leq j \leq m$). The multiplication rules are:

\begin{equation}
1^{'}\times 1^{'} = 1 ; ~~~1^{'}\times 2_{(j)} = 2_{(j)}
\end{equation}
\begin{equation}
2_{(i)}\times 2_{(j)} = \delta_{ij} (1 + 1^{'}) + 2_{(min[i+j,N-i-j])}
+ (1 - \delta_{ij}) 2_{(|i - j|)}
\end{equation}
\noindent

For even N, $D_N$ has four singlets $1, 1^{'},1^{''},1^{'''}$ and $(m - 1)$
doublets
$2_{(j)}$ ($ 1 \leq j \leq m - 1$) where $m = N/2$ with multiplication rules:

\begin{equation}
1^{'}\times 1^{'} = 1^{''} \times 1^{''} = 1^{'''} \times 1^{'''} = 1
\end{equation}
\begin{equation}
1^{'} \times 1^{''} = 1^{'''}; 1^{''} \times 1^{'''} = 1^{'}; 1^{'''} \times
1^{'} = 1^{''}
\end{equation}
\begin{equation}
1^{'}\times 2_{(j)} = 2_{(j)}
\end{equation}
\begin{equation}
1^{''}\times 2_{(j)} = 1^{'''} \times 2_{(j)} = 2_{(m-j)}
\end{equation}
\begin{equation}
2_{(j)} \times 2_{(k)} = 2_{|j-k|} + 2_{(min[j+k,N-j-k])}
\end{equation}

\noindent
(if $k \neq j, (m - j)$)

\begin{equation}
2_{(j)} \times 2_{(j)} = 2 _{(min[2j,N-2j])} + 1 + 1^{'}
\end{equation}

\noindent
(if $j \neq m/2$ )

\begin{equation}
2_{(j)} \times 2_{(m - j)} = 2_{|m - 2j|} + 1^{''} + 1^{'''}
\end{equation}

\noindent
(if $j \neq m/2 $)

\begin{equation}
2_{m/2} \times 2_{m/2} = 1 + 1^{'} + 1^{''} + 1^{'''}
\end{equation}

\noindent
This last is possible only if m is even and hence if N is divisible by {\it
four}.\\

For $Q_{2N}$, there are four singlets $1$,$1^{'}$,$1^{''}$,$1^{'''}$ and
$(N - 1)$ doublets $2_{(j)}$ ($ 1 \leq j \leq (N-1) $).

The singlets have the multiplication rules:

\begin{equation}
1 \times 1 = 1^{'} \times 1^{'} = 1
\end{equation}
\begin{equation}
1^{''} \times 1^{''} = 1^{'''} \times 1^{'''} = 1^{'}
\end{equation}
\begin{equation}
 1^{'} \times 1^{''} = 1^{'''} ; 1^{'''} \times 1^{'} = 1^{''}
\end{equation}

\noindent
for $N = (2k + 1)$ but are identical to those for $D_N$ when N = 2k.

The products involving the $2_{(j)}$ are identical to those given
for $D_N$ (N even) above.

This completes the multiplication rules for 19 of the 45 groups. 
The complete multiplication tables for all the nonabelian groups
with order $g \leq 31$ are provided in Appendix A of \cite{FK3}.

\bigskip
\bigskip
\bigskip
\bigskip

\underline{Mathematical Theorem: A Pseudoreal $4$ of $SU(4)$ Cannot Yield Chiral Fermions.}

\bigskip

In \cite{CV} it was proved that if the embedding in $SU(4)$ is such that
the ${\bf 4}$ is real, the
resultant fermions are always non-chiral. It was implied there that the
converse holds, that if ${\bf 4}$ is complex, ${\bf 4}={\bf 4}^{{\bf *}}$ ,
then the resulting fermions are necessarily chiral. Actually for $\Gamma $ 
$\subset $ $SU(2)$ one encounters the intermediate possibility that the {\bf 4
} is {\it pseudoreal}. In the present section we shall show that if ${\bf 4}$
is pseudoreal then the resultant fermions are necessarily non-chiral. The
converse now holds: if the ${\bf 4}$ is neither real nor pseudoreal then the
resultant fermions are chiral.

\bigskip

For $\Gamma \subset SU(2)$ it is important that the embedding be consistent
with the chain $\Gamma \subset SU(2)\subset SU(4)$, otherwise the embedding
is not a consistent one. One way to see the inconsistency is to check the
reality of the ${\bf 6}=({\bf 4}\otimes {\bf 4)}_{antisymmetric}$. If ${\bf 6
}\neq {\bf 6}^{{\bf *}}${\bf \ }then the embedding is clearly improper. To
avoid this inconsistency it is sufficient to include in the ${\bf 4}$ of 
$SU(4)$ only complete irreducible representations of $SU(2)$.

\bigskip

An explicit example will best illustrate this propriety constraint on
embeddings. Let us consider $\Gamma =Q_{6}$, the dicyclic group of order 
$g=12$. This group has six inequivalent irreducible representations: 
$1,1^{\prime },1^{\prime \prime },1^{\prime \prime \prime },2_{1},2_{2}$. The
1, $1^{\prime }$, 2$_{1}$ are real. The $1^{\prime \prime }$ and $1^{\prime
\prime \prime }$ are a complex conjugate pair, The $2_{2}$ is pseudoreal. To
embed $\Gamma =Q_{6}\subset SU(4)$ we must choose from the special
combinations which are complete irreducible representations of $SU(2)$
namely 1, $2=2_{2}$, $3=1^{\prime }+2_{1}$ and $4=1^{\prime \prime
}+1^{\prime \prime \prime }+2_{2}$. In this way the embedding either makes
the ${\bf 4}$ of $SU(4)$ real {\it e.g}. $4=1+1^{\prime }+2_{1}\ $and the
theorem of \cite{CV} applies, and non-chirality results, or the ${\bf 4}$
is pseudoreal {\it e.g}. $4=2_{2}+2_{2}$. In this case one can check that
the embedding is consistent because $({\bf 4}\otimes {\bf 4)}
_{antisymmetric} $ is real. But it is equally easy to check that the product
of this pseudoreal ${\bf 4}$ with the complete set of irreducible
representations of $Q_{6}$ is again real and that the resultant fermions are
non-chiral.

The lesson is:

{\it To obtain chiral fermions from compactification on }${\it AdS}_{{\it 5}
}\times S_{5}/\Gamma ${\it , the embedding of }${\it \Gamma }${\it \ in }
$SU(4)${\it \ must be such that the 4 of }$SU(4)${\it \ is neither real nor
pseudoreal}.
\bigskip

\bigskip
\bigskip
\bigskip

\noindent Now we are ready for a successful example\cite{FK2,FK3}

\bigskip
\bigskip

\noindent \underline{Group 24/7; also designated $D_4 \times Z_3$}

\bigskip
\bigskip

\noindent This has twelve singlets $1_1\alpha^i, 1_2\alpha^i, 1_3\alpha^i, 1_4\alpha^i$ 
(i = 0 - 2)
and three doublets $2\alpha^i$ (i = 0 - 2); here $\alpha = exp (i\pi/3)$.
The embedding ${\bf 4} = (1_1\alpha, 1_2, 2\alpha)$
was studied in detail in a previous article 
\cite{FK2} where it was shown how it can lead to precisely three
chiral families in the standard model.
For completeness we include the table\cite{FK3} for the chiral fermions (it was
presented in a different equivalent way in \cite{FK2}):

\bigskip
\bigskip

\begin{tabular}{||c||c|c|c|c|c||c|c|c|c|c||c|c|c|c|c||}
\hline
 & $1_1$ & $1_2$ & $1_3$ & $1_4$ & $2$ &$1_1\alpha$& $1_2\alpha$ & $1_3\alpha$ & $1_4\alpha$ & $2\alpha$ 
& $1_1\alpha^2$ & $1_2\alpha^2$ & $1_3\alpha^2$ & $1_4\alpha^2$& $2\alpha^2$  \\
\hline\hline
$1_1$&&$\times$&&&&$\times$&&&&$\times$&&&&& \\
\hline
$1_2$&$\times$&&&&&&$\times$&&&$\times$&&&&& \\
\hline
$1_3$&&&&$\times$&&&&$\times$&&$\times$&&&&& \\
\hline
$1_4$&&&$\times$&&&&&&$\times$&$\times$&&&&& \\
\hline
$2$&&&&&$\times$&$\times$&$\times$&$\times$&$\times$&$\times$&&&&& \\
\hline\hline
$1_1\alpha$&&&&&&&$\times$&&&&$\times$&&&&$\times$ \\
\hline
$1_2\alpha$&&&&&&$\times$&&&&&&$\times$&&&$\times$  \\
\hline
$1_3\alpha$&&&&&&&&&$\times$&&&&$\times$&&$\times$ \\
\hline
$1_4\alpha$ &&&&&&&&$\times$&&&&&&$\times$&$\times$ \\
\hline
$2\alpha$ &&&&&&&&&&$\times$&$\times$&$\times$&$\times$&$\times$&$\times$ \\
\hline\hline
$1_1\alpha^2$&$\times$&&&&$\times$&&&&&&&$\times$&&&  \\
\hline
$1_2\alpha^2$&&$\times$&&&$\times$&&&&&&$\times$&&&&\\
\hline
$1_3\alpha^2$&&&$\times$&&$\times$&&&&&&&&&$\times$& \\
\hline
$1_4\alpha^2$&&&&$\times$&$\times$&&&&&&&&$\times$&&\\
\hline
$2\alpha^2$&$\times$&$\times$&$\times$&$\times$&$\times$&&&&&&&&&&$\times$  \\
\hline\hline
\end{tabular}

\bigskip
\bigskip

\bigskip
\bigskip

\noindent By identifying $SU(4)$ with the diagonal subgroup of
$SU(4)_{2,3}$, breaking $SU(4)_1$ to $SU(2)_L^{'}\times
SU(2)_R^{'}$, then identifying
$SU(2)_L$ with the diagonal subgroup
of $SU(2)_{6,7,8}$ and $SU(2)_L^{'}$
and $SU(2)_R$ with the diagonal subgroup of
$SU(2)_{10,11,12}$ and $SU(2)_R^{'}$
then leads to a three-family model.

\bigskip

\noindent This model is especially interesting because, uniquely among
the large number of models examined in this study, the prescribed scalars are 
sufficient to break the gauge symmetry to that of the standard model
with three chiral families.

\bigskip
\bigskip

\newpage

\bigskip
\bigskip

\section{Gauge Coupling Unification}

\bigskip

Most of the research beyond the standard model\cite{Dine}
is motivated by the hierarchy problem
and uses the two assumptions of grand
unification and low-energy ($\sim TeV$) supersymmetry.
This is, in turn, driven largely by the successful prediction
of one number, $sin^2 \theta$ of the
electroweak mixing angle $\theta$. It is proposed to
replace the two assumptions of grand unification and low-energy
supersymmetry by one assumption, conformality. It therefore
is important to show that $sin^2\theta$
can be derived from conformality alone; that is the
principal objective of this section.

Before entering into conformality, let us
briefly review the alternative. The experimental data give
couplings at the Z pole of\cite{PDG}
$\alpha_3=0.118\pm0.003, \alpha_2=0.0338, \alpha_1=\frac{5}{3}\alpha_Y^{'}
=0.0169$ (where the errors on $\alpha_{1,2}$ are less than 1\%) and
$sin^2\theta = \alpha^{'}_Y /(\alpha_2 + \alpha^{'}_Y) = 0.231$
with an error less than 0.001. Note that $\alpha_2/\alpha_1$ is very nearly two; this
will be used later. The RGE for the supersymmetric
grand unification\cite{S,DG} are
\begin{equation}
\frac{1}{\alpha_i(M_G)} = \frac{1}{\alpha_i(M_Z)} - \frac{b_i}{2 \pi}
ln \left(\frac{M_G}{M_Z} \right)
\label{RGE}
\end{equation}
Using the MSSM values $b_i = (6\frac{3}{5}, 1, -3)$ and substituting
$\alpha_{2,3}$ at $M_Z = 91.187$GeV gives $M_G=2.4\times 10^{16}$GeV
and $\alpha_{2,3}(M_G)^{-1}=24.305$.
Using Eq(\ref{RGE}) with $i=1$ now predicts $\alpha_1(M_Z)
=59.172$ and hence $sin^2\theta=0.231$; this is
impressive agreement with experiment and is sometimes
presented as the accurate meeting of three straight lines on
a $\alpha^{-1}_i(\mu)$ vs. ln$\mu$
plot\cite{ADF,ADFFL}.

As we have seen above,
the relationship of the Type IIB superstring to conformal gauge theory
in $d=4$ gives rise to an interesting class of gauge
theories.
Choosing the simplest compactification\cite{maldacena}
on $AdS_5 \times S_5$ gives rise to an ${\cal N} = 4$ SU(N) gauge theory
which is known to be conformal due to
the extended global supersymmetry and non-renormalization theorems. All
of the RGE $\beta-$functions for this ${\cal N} = 4$
case are vanishing in perturbation theory. It is possible to break
the ${\cal N}=4$ to ${\cal N}=2,1,0$ by replacing
$S_5$ by an orbifold $S_5/\Gamma$
where $\Gamma$ is a discrete group with
$\Gamma \subset SU(2), \subset SU(3), \not\subset SU(3)$
respectively.

In building a conformal gauge theory model \cite{F1,WS,CV},
the steps are: (1) Choose the discrete group $\Gamma$; (2) Embed
$\Gamma \subset SU(4)$; (3) Choose the $N$ of $SU(N)$; and
(4) Embed the Standard Model $SU(3) \times SU(2) \times U(1)$
in the resultant gauge group $\bigotimes SU(N)^p$ (quiver
node identification). Here we shall look only
at abelian $\Gamma = Z_p$ and define
$\alpha = exp(2 \pi i/p)$. It is expected from the string-field
duality that the resultant field
theory is conformal in the $N\longrightarrow \infty$ limit,
and will have a fixed manifold, or at least a fixed point, for $N$ finite.

Before focusing on ${\cal N}=0$ non-supersymmetric cases, let
us first examine an ${\cal N}=1$ model first
put forward in \cite{KS}.
The choice is $\Gamma = Z_3$ and the {\bf 4} of $SU(4)$
is {\bf 4} = $(1, \alpha, \alpha, \alpha^2)$. Choosing N=3,
this leads to the three chiral families under $SU(3)^3$
trinification\cite{DGG}
\begin{equation}
(3, \bar{3}, 1) + (1, 3, \bar{3}) + (\bar{3}, 1, 3)
\end{equation}
In this model it is interesting that
the number of families arises as 4-1=3,
the difference between the 4 of SU(4) and ${\cal N}=1$,
the number of unbroken supersymmetries.
However this model has no gauge coupling unification;
also, keeping ${\cal N}=1$ supersymmetry is against the
spirit of the conformality approach. We now
present three examples, Models A, B, and C, which accommodate
three chiral families, break all supersymmetries
(${\cal N}=0$), and possess gauge coupling unification, including the
correct value of the electroweak mixing angle.

{\it Model A}. Choose $\Gamma = Z_7$, embed the 4 of SU(4)
as $(\alpha^2, \alpha^2, \alpha^{-3}, \alpha^{-1})$, and
choose N=3 to aim at a trinification
$SU(3)_C \times SU(3)_W \times SU(3)_H$.

The seven nodes of the quiver diagram will
be identified as C-H-W-H-H-H-W.

The behavior of the 4 of SU(4) implies that the bifundamentals
of chiral fermions are in the representations
\begin{equation}
\sum_{j=1}^{7} [ 2(N_j, \bar{N}_{j+2}) + (N_j, \bar{N}_{j-3})
+ (N_j, \bar{N}_{j-1})]
\end{equation}
Embedding the C, W and H SU(3) gauge groups as
indicated by the quiver mode identifications
then gives the seven quartets of irreducible representations:
\begin{equation}
\begin{array}{l}
[3(3,\bar{3}, 1) + (3, 1, \bar{3})]_1 + \\

+ [3(1, 1, 1+8) + (\bar{3}, 1, 3)]_2 + \\

+ [3(1, 3, \bar{3}) + (1, 1 + 8, 1)]_3 + \\

+ [(2(1, 1, 1+8) + (1, \bar{3}, 3) + (\bar{3}, 1, 3)]_4 + \\

+ [2(1, 1, 1 + 8) + 2 (1, \bar{3}, 3)]_5 + \\

+ [2(\bar{3}, 1, 3) + (1, 1, 1 + 8) + (1, \bar{3}, 3)]_6 + \\

+ [4(1, 3, \bar{3})]_7.
\end{array}
\end{equation}
Combining terms gives, aside from (real) adjoints and overall singlets
\begin{equation}
3(3, \bar{3}, 1) + 4(\bar{3}, 1, 3) + (3, 1, \bar{3})
+ 7(1, 3, \bar{3}) + 4(1, \bar{3}, 3)
\end{equation}
Cancelling the real parts (which acquire Dirac masses at the conformal
symmetry breaking scale) leaves under trinification
$SU(3)_C \times SU(3)_W \times SU(3)_H$
\begin{equation}
3 [(3, \bar{3}, 1) + (1, 3, \bar{3}) + (\bar{3}, 1, 3)]
\end{equation}
which are the desired three chiral families.

Given the embedding of $\Gamma$ in SU(4) it follows that the 6 of SU(4) transforms
as
$(\alpha^4, \alpha, \alpha, \alpha^{-1}, \alpha^{-1}, \alpha^{-4})$.
The complex scalars therefore transform as
\begin{equation}
\sum_{j=1}^{7} [(N_j, \bar{N}_{j\pm4}) + 2 (N_j, \bar{N}_{j\pm1})]
\end{equation}
These bifundamentals can by their VEVS break the symmetry $SU(3)^7$
= $SU(3)_C \times SU(3)_W^2 \times SU(3)^4_H$ down
to the appropriate diagonal subgroup
$SU(3)_C \times SU(3)_W \times SU(3)_H$.

\bigskip
\bigskip
\bigskip

{\it Now to the final aspect of Model A which is its motivation, the gauge coupling
unification. The embedding in $SU(3)^7$ of
$SU(3)_C \times SU(3)^2_W \times SU(3)^4_H$ means that the
couplings $\alpha_1, \alpha_2, \alpha_3$ are
in the ratio
$\alpha_1 / \alpha_2 / \alpha_3 = 1/2/4$.
Using the phenomenological data given at the beginning,
this implies that
$sin^2\theta = 0.231$}. 

\bigskip
\bigskip

On the other hand, the QCD coupling is
$\alpha_3 = 0.0676$ which is too low unless the
conformal scale is at least 10TeV.
We prefer a scale $\sim 1$ TeV
for conformal breaking where
$\alpha_3$ is nearer to 0.10. This motivates our Models B and C below
which have larger $\alpha_3$ but are otherwise
more complicated.

\bigskip

{\it Model B.} Choose $\Gamma = Z_{10}$  and embed $Z_{10}\subset SU(4)$
such that 4 = $(\alpha^4, \alpha^4, \alpha^{-3}, \alpha^{-5})$.
The chiral fermions are therefore
\begin{equation}
\sum_{j=1}^{10} [2(N_j, \bar{N}_{j+4}) + (N_j, \bar{N}_{j-3})
+ (N_j, \bar{N}_{j-5})]
\end{equation}
To attain trinification we identify the quiver nodes as
C-H-H-H-W-W-H-W-H-H and then the chiral fermions are in the ten
quartets of irreducible representations
\begin{equation}
\begin{array}{l}
[4(3, \bar{3}, 1)]_1 + \\

+ [2(1, \bar{3}, 3) + (1, 1, 1+8)]_2 + \\

+ [2(1, 1, 1+8) + (1, \bar{3}, 3)]_3 + \\

+ [2(1, \bar{3}, 3) + (\bar{3}, 1, 3) + (1, 1, 1+8)]_4  + \\

+ [4(1, 3, \bar{3})]_5 + \\

+ [ 3(1, 3, \bar{3}) + (\bar{3}, 3, 1)]_6 + \\

+ [2(\bar{3}, 1, 3) + (1, 1, 1+8)]_7 + \\

+ [3(1, 3, \bar{3}) + (1, 1+8, 1)]_8 + \\

+ [3(1, 1, 1+8) + (1, \bar{3}, 3)]_9 + \\

+ [3( 1, 1, 1+8) + (1, \bar{3}, 3)]_{10}
\end{array}
\end{equation}
Removing the (real) octets and singlets leaves
\begin{equation}
4(3, \bar{3}, 1) + (\bar{3}, 3, 1) + 3(\bar{3}, 1, 3) + 10(1, 3, \bar{3}) + 7(1, \bar{3}, 3)
\end{equation}
so that the chiral (complex) part is again
\begin{equation}
3[(3, \bar{3}, 1) + (1, 3, \bar{3}) + (\bar{3}, 1, 3)]
\end{equation}
which are three chiral families.

The 6 of SU(4) transforms under $\Gamma = Z_{10}$ as
6 = $(\alpha^8, \alpha, \alpha, \alpha^{-1}, \alpha^{-1}, \alpha^{-8})$
and so the complex scalars are
\begin{equation}
\sum_{j=1}^{10} [(n_j, \bar{N}_{j\pm8}) + 2(N_j, \bar{N}_{j\pm1})]
\end{equation}
With the given quiver node identification VEVs for these scalars can
break $SU(3)^{10} = SU(3)_C \times SU(3)_W^3 \times SU(3)_H^6$ to
the diagonal subgroup $SU(3)_C \times SU(3)_W \times SU(3)_H$.

The couplings $\alpha_1, \alpha_2, \alpha_3$ are in the ratio
$\alpha_1 / \alpha_2 / \alpha_3 = 1/2/6$ corresponding to
$sin^2\theta = 0.231$ and $\alpha_3 = 0.101$. This is within the range
of a TeV conformal breaking scale. Nevertheless, it
is numerically irresistible to notice that the Z-pole
values satisfy $\alpha_1 / \alpha_2 / \alpha_3 = 1/2/7$ which
leads naturally to Model C.

\bigskip

{\it Model C.} Choose $\Gamma=Z_{23}$ and embed in SU(4)
by 4 = $(\alpha^6, \alpha^6, \alpha^{-5}, \alpha^{-7})$.
Given this embedding the quiver nodes can be chosen as
C-C-X-X-X-H-H-W-H-X-X-X-X-X-X-X-W-H-H-W-X-X-X
where the thirteen X's denote any distribution of
four W's and nine H's
that allows breaking by the complex scalars
cited below. The quiver is arranged such
that according to the rule of ($3_C-\bar{3}_W$) minus
($3_W-\bar{3}_C$) there
are three chiral families.
[The model in \cite{CV} did not follow this
rule and has two families.]
Note that because
of anomaly cancellation and the occurrence of only
bifundamentals the remainder of trinification
is automatic and need not be checked in every case.

The chiral families are as in Models A and B.

The 6 of SU(4) transforms as
$(\alpha^{12}, \alpha, \alpha,
\alpha^{-1}, \alpha^{-1}, \alpha^{-12})$.
This implies complex scalars whose VEVs can break
$SU(3)^{23}= SU(3)_C^2 \times SU(3)_W^7 \times SU(3)^{14}_W$
to $SU(3)_C \times SU(3)_W \times SU(3)_H$ with
a suitable distribution of W and H nodes on the quiver.

With this choice of diagonal subgroups the couplings
are in the ratio $\alpha_1 / \alpha_2 / \alpha_3 =
1/2/7$ corresponding to $sin^2\theta = 0.231$ and
$\alpha_3 = 0.118$ which coincide with the Z-pole values.

\bigskip

In this section, we have given three examples of
building conformal models
from abelian $\Gamma$ with acceptable values of the
couplings at the conformal scale, assuming
that the SU(3) gauge couplings are all equal at the conformal
scale. Model A is the simplest but its $\alpha_3$ is too small unless the
conformal scale is taken up to at least 10TeV. Models B and C
can accommodate a lower conformal scale but are more
complicated.

\bigskip

There are two features of conformal models which bear repetition:

(1) Bifundamentals prohibit
representations like (8,2) or (3,3) in the Standard Model consistent
with Nature.

(2) Charge quantization is incorporated since the abelian
$U(1)_Y$ group has a positive-definite $\beta-$function
and cannot be conformal until it is embedded
in a non-abelian group.

\bigskip

There are three questions which merit further investigation:

(1) The first question bears on whether there is a
fixed manifold (line, plane,...) with
respect to the renormalization group
or only a fixed point which is, in any case,
sufficient to apply our conformality constraints.
In perturbation theory, do the $\beta-$functions vanish?

(2) Are the additional particles necessary to render
the Standard Model
conformal consistent with the stringent
constraints imposed by the
precision electroweak data?

(3) Coefficients of dimension-4 operators are prescribed by group
theory and all dimensionless properties such
as quark and lepton mass ratios and mixing angles
are calculable. Do these work and, if not, can one
refine the model-building to obtain a best fit?

\bigskip
\bigskip

\newpage

\bigskip
\bigskip

\section{Discussion}

\bigskip
\bigskip

String theory has existed for over thirty years and its connection with the real
world (at least for ten/eleven dimensional versions) is unknown despite a multitude
of attempts. The AdS/CFT correspondence offers, in our opinion, the
most promising approach presently available to
relate string theory to observable physics.

The use of the AdS/CFT 
correspondence involves the step of, in
the first order, dropping the gravitational interaction.
In any foreseeable high-energy experiment 
gravity will be negligible so the approximation is reasonable.
On the other hand, if the particles predicted by the conformality approach discussed
in this article were to be discovered in the next (TeV) energy regime,
it would provide support for the string approach, including as a theory of quantum gravity.

Whether or not string theory is the correct unifying theory with gravity, 
it does provide through the AdS/CFT correspondence very promising
ideas of how to write conformal
theories in four space-time dimensions
with particular semi-simple gauge groups, chiral fermions and complex scalars
and one of these theories could be the correct direction to
proceed.

\newpage

\bigskip
\bigskip

\section*{Acknowledgements}

This work
was supported in part by the US Department of Energy
under Grant No. DE-FG02-97ER-41036.
I thank W.F. Shively, C. Vafa and T.W. Kephart for collaboration
at different stages of this ongoing project.

\end{document}